\documentclass[twocolumn,prl,aps,amsfonts,showpacs]{revtex4}

\usepackage{graphicx}
\usepackage{amssymb}

\begin{document}
%\draft
\preprint{}

\newcommand{\1}{{\bf \scriptstyle 1}\!\!{1}}
\newcommand{\I}{{\rm i}}
\newcommand{\p}{\partial}
\newcommand{\D}{^{\dagger}}
\newcommand{\bx}{{\bf x}}
\newcommand{\br}{{\bf r}}
\newcommand{\bk}{{\bf k}}
\newcommand{\bv}{{\bf v}}
\newcommand{\bp}{{\bf p}}
\newcommand{\bs}{{\bf s}}
\newcommand{\bu}{{\bf u}}
\newcommand{\bA}{{\bf A}}
\newcommand{\bB}{{\bf B}}
\newcommand{\bE}{{\bf E}}
\newcommand{\bF}{{\bf F}}
\newcommand{\bI}{{\bf I}}
\newcommand{\bK}{{\bf K}}
\newcommand{\bL}{{\bf L}}
\newcommand{\bP}{{\bf P}}
\newcommand{\bQ}{{\bf Q}}
\newcommand{\bS}{{\bf S}}
\newcommand{\bH}{{\bf H}}
\newcommand{\balpha}{\mbox{\boldmath $\alpha$}}
\newcommand{\bsigma}{\mbox{\boldmath $\sigma$}}
\newcommand{\bSigma}{\mbox{\boldmath $\Sigma$}}
\newcommand{\bOmega}{\mbox{\boldmath $\Omega$}}
\newcommand{\bpi}{\mbox{\boldmath $\pi$}}
\newcommand{\bphi}{\mbox{\boldmath $\phi$}}
\newcommand{\bnabla}{\mbox{\boldmath $\nabla$}}
\newcommand{\bmu}{\mbox{\boldmath $\mu$}}
\newcommand{\bepsilon}{\mbox{\boldmath $\epsilon$}}

\newcommand{\iLambda}{{\it \Lambda}}
\newcommand{\cA}{{\cal A}}
\newcommand{\cD}{{\cal D}}
\newcommand{\cF}{{\cal F}}
\newcommand{\cL}{{\cal L}}
\newcommand{\cH}{{\cal H}}
\newcommand{\cI}{{\cal I}}
\newcommand{\cM}{{\cal M}}
\newcommand{\cO}{{\cal O}}
\newcommand{\cR}{{\cal R}}
\newcommand{\cU}{{\cal U}}
\newcommand{\cT}{{\cal T}}

\newcommand{\be}{\begin{equation}}
\newcommand{\ee}{\end{equation}}
\newcommand{\bea}{\begin{eqnarray}}
\newcommand{\eea}{\end{eqnarray}}
\newcommand{\beqa}{\begin{eqnarray*}}
\newcommand{\eeqa}{\end{eqnarray*}}
\newcommand{\nn}{\nonumber}
\newcommand{\DD}{\displaystyle}

\newcommand{\ba}{\left[\begin{array}{c}}
\newcommand{\baa}{\left[\begin{array}{cc}}
\newcommand{\baaa}{\left[\begin{array}{ccc}}
\newcommand{\baaaa}{\left[\begin{array}{cccc}}
\newcommand{\ea}{\end{array}\right]}

%%%%%%%%%%%%%%%%%%%%%%%%%%%%%%%%%%%%%%%%%%%%%%%%%%%%%%%%%%%%%%%%%%%%%%%%%%%%%%

\title{Berry-phase oscillations of the Kondo effect in single-molecule magnets}

\author{Michael N.~Leuenberger$^{1,2}$}\email{mleuenbe@mail.ucf.edu}
\author{Eduardo R. Mucciolo$^2$}\email{mucciolo@physics.ucf.edu}

\affiliation{$^1$NanoScience Technology Center, University of Central
Florida, Orlando, FL 32826} 

\affiliation{$^2$Department of Physics, University of Central
Florida, Orlando, FL 32816}

%%%%%%%%%%%%%%%%%%%%%%%%%%%%%%%%%%%%%%%%%%%%%%%%%%%%%%%%%%%%%%%%%%%%%%%%%%%%%%

\date{\today}

\begin{abstract}
We show that it is possible to topologically induce or quench the
Kondo resonance in the conductance of a single-molecule magnet
($S>1/2$) strongly coupled to metallic leads. This can be achieved by
applying a magnetic field perpendicular to the molecule easy axis and
works for both full- and half-integer spin cases. The effect is caused
by the Berry-phase interference between two quantum tunneling paths of
the molecule's spin. We have calculated the renormalized Berry-phase
oscillations of the Kondo peaks as a function of the transverse
magnetic field as well as the conductance of the molecule by means of
the poor man's scaling method. We propose to use a new variety of the
single-molecule magnet Ni$_4$ for the experimental observation of this
phenomenon.
\end{abstract}

\pacs{72.10.Fk, 03.65.Vf, 75.45.+j, 75.50.Xx}

\maketitle

%%%%%%%%%%%%%%%%%%%%%%%%%%%%%%%%%%%%%%%%%%%%%%%%%%%%%%%%%%%%%%%%%%%%%%%%%%%%%%

The quantum tunneling of the spin of single-molecule magnets (SMMs),
such as Mn$_{12}$ \cite{experiments,delBarco} and Fe$_8$
\cite{Sangregorio,Wernsdorfer}, has attracted a great deal of
interest. These molecules have a large total spin, strong uniaxial
anisotropy, and interact very weakly when forming a crystal. They have
already been proposed for high-density magnetic storage as well as
quantum computing applications \cite{app}. Yet, there is much to
explore in their fundamental properties. For instance, recent
measurements of the magnetization in bulk Fe$_8$ samples (see
Ref.~\cite{Zener}) have observed oscillations in the tunnel splitting
$\Delta E_{m,m'}$ between states $S_z = m$ and $m'$ as a function of a
transverse magnetic field at temperatures between $0.05$ K and $=0.7$
K. Using a coherent spin-state path integral approach, it has been
shown that this effect results from the interference between Berry
phases carried by spin tunneling paths of opposite windings
\cite{Loss,Delft,Garg}, a concept also applicable to transitions
involving excited states of SMMs \cite{Leuenberger_Berry}.

A new approach to the study of SMMs opened up recently with the first
observations of quantized electronic transport through an {\it
isolated} Mn$_{12}$ molecule \cite{newexps}. One expects a rich
interplay between quantum tunneling, phase coherence, and electronic
correlations in the transport properties of SMMs. It has been argued
that the Kondo effect would only be observable for SMMs with
half-integer spin \cite{Romeike_Kondo} and therefore absent for SMMs
such as Mn$_{12}$, Fe$_8$, and Ni$_4$, where the spin is integer. Here
we show that this prediction is only valid in the absence of an
external magnetic field. Remarkably, even a moderate transverse
magnetic field topologically quenches the two lowest levels of a
full-integer spin SMM, making them degenerate. The same Berry-phase
interference also affects transport for SMMs with half-integer spin:
In that case, sweeping the magnetic field will lead not to one but
{\it a series} of Kondo resonances.

It is interesting to contrast the Kondo effect in a SMM with that
observable in a lateral quantum dot with a single excess electron
\cite{Goldhaber,Cronenwett}, in a single spin-1/2 atom \cite{Park}, or
in a single spin-1/2 molecule \cite{Liang}. In those cases, at zero
bias the Kondo effect is damped by an external a magnetic field
because the degeneracy of the two spin states is lifted
\cite{Glazman2005}. In the case of SMMs, the Berry-phase oscillations
of the tunnel splitting $\Delta E_{m,m'}$ leads to oscillation of the
Kondo effect as a function of $H_\bot$, the transverse magnetic field
amplitude. This means that the Kondo effect is observable at zero bias
for all values of $H_{\bot,0}$ such that $\Delta
E_{m,m'}(H_{\bot,0})=0$. Notice that at a finite bias the Kondo effect
in a quantum dot in the presence of a magnetic field of magnitude $H$
can be restored by tuning the bias to $eV = \pm g \mu_B H$ (see
Ref. \cite{Goldhaber}). For SMMs, however, the interference between
the Berry phases accumulated by the molecule's spin makes the distance
between the split Kondo peaks, which is equal to $eV = \pm \Delta
E_{m,m'}$, oscillate as a function of $H_\bot$.

%In the following, we present our calculations and discuss our results
%in detail. We combine a standard model of a SMM coupled to metallic
%electrodes through tunnel barriers with Anderson's poor man's scaling
%description of the Kondo problem. We derive expressions for the
%conductance at zero and finite bias as a function of a transverse
%magnetic field and comment on their experimental significance.

%%%%%%%%%%%%%%%%%%%%%%%%%%%%%%%%%%%%%%%%%%%%%%%%%%%%%%%%%%%%%%%%%%%%%%%%%%%%%%

Consider a typical spin Hamiltonian of a SMM in an external transverse
magnetic field $H_\bot$:
\begin{eqnarray}
\cH_{\rm spin} & = & -A_qS_{q,z}^2+\frac{B_q}{2} \left( S_{q,+}^2
+ S_{q,-}^2 \right) + \frac{B_{4,q}}{3} \left( S_{q,+}^4 + S_{q,-}^4
\right) \nonumber \\ & & + \frac{1}{2}(h_\bot^* S_{q,+}+h_\bot
S_{q,-}),
\label{H_spin}
\end{eqnarray}
where the easy axis is taken along $z$, $S_{q,\pm} = S_{q,x} \pm
iS_{q,y}$, the integer index $q$ denotes the charging state of the
SMM, and $h_\bot=g\mu_B H_\bot$. Note that the transverse magnetic
field lies in the $xy$ plane. In this Hamiltonian, the dominant
longitudinal anisotropy term creates a ladder structure in the
molecule spectrum where the $|\pm m_q\rangle$ eigenstates of $S_z$ are
degenerate. The weak transverse anisotropy terms couple these
states. The coupling parameters depend on the charging state of the
molecule. For example, it is known that Mn$_{12}$ changes its
easy-axis anisotropy constant (and its total spin) from $A_0=56$
$\mu$eV ($S_0=10$) to $A_{-1}=43$ $\mu$eV ($S_{-1}=19/2$) and
$A_{-2}=32$ $\mu$eV ($S_{-2}=10$) when singly and doubly charged,
respectively \cite{dataexps}.

The spin tunneling between the states $\left|m_q\right>$ to
$\left|-m_q\right>$, with $|m_q|\leq S_q$, can occur both clockwise
and counterclockwise around the $x$ axis. 
%
%Let us call $H_x = |H_\bot|\cos\phi$. If $B_{4,q}=0$ (applicable to
%Mn$_{12}$ and Fe$_8$), the paths acquire one of the two Berry phases
%\cite{Garg,Leuenberger_Berry}
%
%$ 
%\Phi_\pm(m_q) = \pm \pi m_q \left[ 1 +
%\frac{h_x}{4m_q\sqrt{B_q(A_q+2B_q)}} \right], 
%$
%
%where $h_x = g\mu_B |H_\bot|\cos\varphi$. The total tunneling
%amplitude is thus proportional to $\cos(\Phi_+)$, leading to Berry
%phase oscillations in $h_x$ with period $4\sqrt{B_q(A_q+2B_q)}$. The
%tunneling amplitude for a full-integer spin is zero at the zero points
%$h_{x,0,n} = (2n+1)\sqrt{4B_q(A_q+2B_q)}$, while for a half-integer
%spin one finds that $h_{x,0,n}=2n\sqrt{4B_q(A_q+2B_q)}$, where $n$ is
%an integer. Since $0\le\left|\Phi_+-\Phi_-\right|\le 4\pi m_q$, there
%are $2m_q$ zeroes of the tunnel splitting $\Delta E_{m_q,-m_q}$.
%
These two paths interfere with each other, which leads to Berry-phase
oscillations \cite{Garg,Leuenberger_Berry}.
Experiments with Ni$_4$
%, however, 
show that $B_{4,q=0}=-0.003$ K, i.e. $B_{4,0}$ is negative. In this
case, in order to see the Berry-phase oscillation, the transverse
magnetic field must be applied along angles that depend on the values
of $B_q$ and $B_{4,q}$ \cite{Leuenberger_Berry}. In Fig. 1 we show the
Berry-phase oscillations of the tunnel splitting calculated for Ni$_4$
for two of such special orientations based on data from
Ref. \cite{Sieber}.

%%%%%%%%%%%%%%%%%%%%%%%%%%%%%%%%%%%%%%%%%%%%%%%%%%%
\begin{figure}[htb]
\includegraphics[width=7cm]{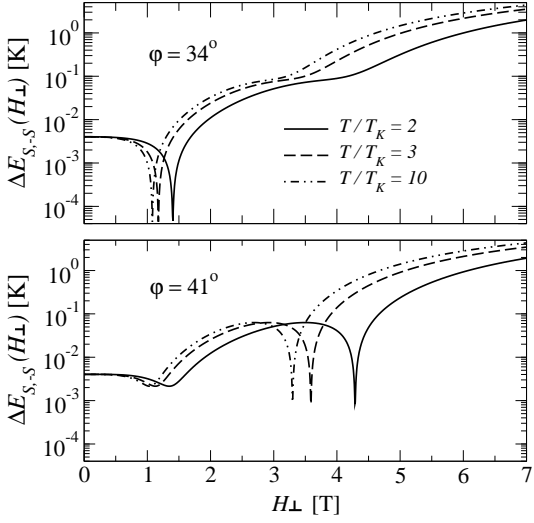}
\caption[]{Berry-phase oscillations of the spin tunnel splitting in
the Ni$_4$ spin cluster of Ref. \cite{Sieber} at different
temperatures ($S=4$). The anisotropy constants are $A=-1.33$ K,
$B=0.034$ K, and $B_4^4=-0.003$ K. $\Delta E_{S,-S}$ is calculated by
an exact diagonalization for $\varphi = 33.5^\circ$ and $\varphi =
40.75^\circ$. As $T$ approaches $T_K$, the renormalization of the
$g$ factor increases, an effect that can be verified
experimentally. The bare $g$ factor is $g=2.2$ for Ni$_4$.}
\label{Berryoscillations}
\end{figure}
%%%%%%%%%%%%%%%%%%%%%%%%%%%%%%%%%%%%%%%%%%%%%%%%%%%

In order to show how these oscillations impact transport through the
SMM, we first evaluate the Kondo effect for zero bias at the zero
points, where the states $\left|m_q\right>$ and $\left|-m_q\right>$
are pairwise degenerate for all $m_q$. In order to consider the Kondo
effect for SMMs, we need to add to the spin Hamiltonian in
Eq. (\ref{H_spin}) the Kondo Hamiltonian \cite{Glazman2005}
\begin{eqnarray} 
\cH_{\rm K} & = & \sum_{k,s} \left[\xi_k + \frac{1}{2}(h_\bot^*s_+
+h_\bot s_-)\right] \psi_{ks}\D \psi_{ks}  +
\sum_{m_q} \left[ J_z^{(m_q)}\right. \nonumber \\ & & \times\left.s_z\Sigma_z^{(m_q)} + \frac{1}{2}
J_\perp^{(m_q)} \left( s_+\Sigma_-^{(m_q)} +
s_-\Sigma_+^{(m_q)}\right)\right],
\label{H_Kondo}
\end{eqnarray}
where $\bs = \sum_{k,k',s,s'} \psi_{ks} \D (\bsigma_{ss'}/2)
\psi_{k's'}$ and the operators $\psi_{ks}^\dagger$ ($\psi_{ks}$)
create (annihilate) electronic states in the leads with momentum $k$,
spin $s$, and energy $\xi_{k}$. $\bSigma^{(m_q)}$ is the pseudospin
1/2 operator acting on the states $\left|m_q\right>$ and
$\left|-m_q\right>$. We define $\Sigma_{\pm}^{(m_q)} =
\Sigma_{x}^{(m_q)} \pm i\Sigma_{y}^{(m_q)} = \left|\pm
m_q\right>\left<\mp m_q\right|$ and $\Sigma_z^{(m_q)} =
\left(\left|m_q\right>\left<m_q\right| - \left|-m_q\right>
\left<-m_q\right| \right)/2$. The spin-flip terms in
Eq. (\ref{H_Kondo}) induce the Kondo resonance. The exchange part,
$\cH_{\rm ex}$ [the second term on the right-hand side of
Eq. (\ref{H_Kondo})] can be derived from a generalized Anderson's
impurity model that takes into account the charging dependence of the
total spin of the SMM. When the charging energy $U$ is much larger
than the tunneling matrix element $t$ between the leads and the SMM,
the exchange coupling can be derived in second-order perturbation
theory by means of a Schrieffer-Wolff transformation
\cite{SchriefferWolff}, yielding
\bea J_\perp^{(m_q)} & = & \frac{2t^2}{U} \left[ \frac{\Delta
E_{m_{q-1},-m_{q-1}}}{\frac{U}{2} + \Delta E_{m_{q-1},-m_{q-1}}} +
\frac{\Delta E_{m_{q+1},-m_{q+1}}}{\frac{U}{2} + \Delta
E_{m_{q+1},-m_{q+1}}}\right] \nn \\ & \approx & 4t^2 \left( \Delta
E_{m_{q-1},-m_{q-1}} + \Delta E_{m_{q+1},-m_{q+1}} \right)/U^2, \nonumber \\
J_z^{(m_q)} & = & \frac{4t^2}{U} \left[ \frac{U+\Delta
E_{m_{q-1},-m_{q-1}}}{\frac{U}{2} + \Delta E_{m_{q-1},-m_{q-1}}}
+\frac{U+\Delta E_{m_{q+1},-m_{q+1}}}{\frac{U}{2} + \Delta
E_{m_{q+1},-m_{q+1}}}\right] \nonumber \\ & \approx & 8t^2/U, \eea
where $U$ is the charging energy. Notice that the coupling is
antiferromagnetic and $J_z^{(m_q)}\gg J_\perp^{(m_q)}$. Since
$J_\perp^{(m_q)}$ depends on the tunnel splittings $\Delta
E_{m_{q-1},-m_{q-1}}$ and $\Delta E_{m_{q+1},-m_{q+1}}$ (which
correspond to one charge added and removed from the SMM,
respectively), the Kondo exchange coupling is strongly anisotropic,
which confirms the result obtained in Ref.~\cite{Romeike_Kondo}.

The transverse anisotropy terms in Eq.~(\ref{H_spin}) mix the states
$\left|m_q\right>$. Since $|A| \gg |B|,|B_4|S^2$, the eigenstates are
nondegenerate and of the form $\left|m_q^{(\pm)}\right> = \left(
\left|m_q\right> \pm \left| - m_q\right> \right)/\sqrt 2$. However, at
the zero points the SU(2) symmetry is restored, yielding degenerate
eigenstates of the form $\left|\pm m_q\right>$. The Kondo Hamiltonian
in Eq.~(\ref{H_Kondo}) opens up transition paths between pairs of
degenerate states $\left|m_q\right>$ and $\left|-m_q\right>$ and these
paths depend on temperature. This leads in general to multi-path (but
single-channel) Kondo correlations at $T>T_K$. At $T=0$, however, only
the $\left|\pm S\right>$ states contribute to the Kondo effect. The
unusual feature of Eq. (\ref{H_Kondo}) is that the total spin is not
conserved. After a spin-flipping event, the excess angular momentum
$L=2m_{q-1}\hbar$ must be absorbed by orbital (and possibly nuclear)
degrees of freedom in the SMM and then be transfered to the molecule
as a whole. Since the kinetic energy of a rotation of tens of $\hbar$
corresponds to a few mK for a typical SMM, the excess orbital angular
momentum will be relaxed by thermal fluctuations in the metallic
contacts. The critical assumption we make is that the excess angular
momentum is transfered from spin to orbital (and possibly nuclear)
degrees of freedom fast enough as to allow for the Kondo state to be
formed.

Our analysis employs the standard poor man's scaling \cite{Anderson}
to renormalize the effective exchange coupling constants $J_z$, $J_+$,
$J_-$, and the $g$ factor. In order to make the discussion
self-contained, we present the main steps of the derivation. We start
by calculating the renormalization flow at the zero points where the
Kondo effect is observable for zero bias. It is reasonable to assume
that $h_{x,0,n}\gg T_K$, except when $n=0$ for half-integer spins. The
total Hamiltonian for the combined SMM and leads system reads (e will
suppress the index $q$ hereafter)
\begin{eqnarray}
\label{eq:Htot}
\cH_{\rm tot} & = & \sum_{m\ge 0} \left[
\epsilon_{m}\left(\Sigma_z^{(m)}\right)^2 + \frac{1}{2}\eta^{(m)}
\left (\tilde{h}_\bot^* \Sigma_+^{(m)} + \tilde{h}_\bot \Sigma_-^{(m)}
\right)\right] \nonumber \\ & & + \sum_{k,s} \xi_k\psi_{ks}\D
\psi_{ks}+\cH_{\rm ex},
\end{eqnarray}
where $\epsilon_{m}$ is the eigenvalue of $\left|m\right>$ and
$\left|-m\right>$ at the zero points, $|\tilde{h}_\bot|$ is the
effective Zeeman splitting between $\left( e^{-i\varphi/2} \left|m
\right> + e^{i\varphi/2} \left|-m\right> \right)/ \sqrt 2$ and
$\left( e^{-i\varphi/2} \left|m\right> - e^{i\varphi/2} \left| -m
\right> \right)/\sqrt 2$, and $\eta^{(m)} = 1 - \rho_0j_\bot^{(m)}/2$
due to the Knight shift, with $\rho_0$ denoting the density of states
of the itinerant electrons. The Zeeman term for the itinerant
electrons is absent in Eq. (\ref{eq:Htot}) because at finite values of
$h_{\bot,0,n}$ one has to cut the edges of the spin-up and spin-down
bands in the leads to make them symmetric with respect to the Fermi
energy \cite{Glazman2005}. Let us call $D$ the resulting band
width. The Hamiltonian remains invariant under renormalization group
transformations. Using a one-loop expansion (second-order perturbation
theory), we obtain the flow equations
\bea 
\frac{dJ_\pm^{(m)}}{d\zeta} & = & -2\rho_0\, J_\pm^{(m)} J_z^{(m)},
\qquad \frac{dJ_z^{(m)}}{d\zeta}=-2\rho_0\, J_+^{(m)}
J_-^{(m)}, \label{j} \\ 
\frac{d\eta^{(m)}}{d\zeta} & = &
\frac{\rho_0^2}{2}\, \left( J_+^{(m)} + J_-^{(m)} \right) J_z^{(m)},
\eea
where $\zeta = \ln(\tilde{D}/D)$ and $\tilde{D}$ is the rescaled band
width. Dividing Eqs.~(\ref{j}) and integrating by parts gives $\left(
J_z^{(m)} \right)^2 - \left( J_\bot^{(m)} \right)^2 = C^{(m)}$, where
$C^{(m)}$ is a positive constant \cite{Anderson}. The exchange
coupling constants always flow to an antiferromagnetic state because
$J_z^{(m)}>0$, but at the end of the flow the coupling tends to become
isotropic. Solving Eqs.~(\ref{j}) yields
\be
\label{eq:flowsol}
\frac{1}{2\rho_0\sqrt{C^{(m)}}}\, {\rm arctanh} \left(
\frac{\sqrt{C^{(m)}}} {J_z^{(m)}} \right) = \ln \left(
\frac{\tilde{D}}{T_K} \right).
\ee
The solution for $J_\bot^{(m)}$ is determined by $J_z^{(m)} =
\sqrt{\left(J_\bot^{(m)}\right)^2+C^{(m)}}$. The flow stops at
$\tilde{D}\approx \omega \sim T > T_K$. The result of the flow for
$\eta$ is presented in Fig. 2.

We are now ready to calculate the linear conductance through the SMM,
which is given by the formula \cite{Glazman2005}
\be 
G = G_0\int d\omega\left( -\frac{df}{d\omega} \right) \frac{\pi^2
\rho_0^2}{16} \frac{\sum_me^{-\epsilon_m/T} \left|
\cA^{(m)}(\omega)\right|^2} {\sum_me^{-\epsilon_m/T}}, 
\label{conductance}
\ee
where $G_0$ is the classical (incoherent) conductance of the
molecule. At the end of the flow the transition amplitude $\cA^{(m)}$
can be calculated in first-order perturbation theory and one finds
that $\eta_{\tilde{D}\approx\omega}^{(m)} = 1 - \rho_0
J_{\bot,\omega}^{(m)}/2$ and
\bea 
\cA_{\tilde{D}\approx\omega}^{(m)} & = & 
J_{\bot,\omega}^{(m)}  =
\sqrt{C^{(m)}} \left[ \frac{(\omega/T_K)^{2\rho_0 \sqrt{C^{(m)}}}}
{(\omega/T_K)^{4\rho_0 \sqrt{C^{(m)}}}-1} \right],
\label{amplitude} 
%\\
%\eta_{\tilde{D}\approx\omega}^{(m)} & = & 1 - \rho_0
%J_{\bot,\omega}^{(m)}/2. 
%\label{eta}
\eea
Substituting $\omega$ by $T$ and inserting Eq. (\ref{amplitude}) into
(\ref{conductance}), one finds that the conductance diverges as $T
\rightarrow T_K$, signaling the onset of the Kondo effect. Since
$C^{(m)}>0$ the singularity in Eq. (\ref{amplitude}) differs from the
usual logarithmic behavior. The Kondo effect causes a fundamental
change in the SMM behavior: All the zero points of the Berry-phase
oscillation get rescaled by the $g$-factor renormalization:
$b_{\bot,0,n}^{(m)} = h_{\bot,0,n}/\eta_{\tilde{D} \approx
T}^{(m)}$. Thus, the zero points become dependent on the contributing
states $\left|m\right>$ and $\left|-m\right>$. This result indicates
that the period of the Berry-phase oscillations becomes temperature
dependent as $T$ is lowered toward $T_K$. Remarkably, the scaling
equations can be checked experimentally by measuring the renormalized
zero points of the Berry phase. Moreover, due to the scale invariance
of the Kondo effect, the period of oscillations should follow a
universal function of $T/T_K$.

The necessary conditions for observing these oscillations are a large
enough tunnel splitting and a strong coupling between the SMM and the
leads. In regard to the former, a new Ni$_4$ single-molecule magnet
with $S=4$ has been synthesized \cite{Sieber} with $\Delta
E_{S,-S}\sim 0.01$ K or larger, depending on $h_\bot$. However, the
two recent reports of transport through a SMM \cite{newexps} show that
the electrical contacts between the SMM and the leads is rather poor
and would need to be improved in order to bring $T_K$ to accessible
values.

%%%%%%%%%%%%%%%%%%%%%%%%%%%%%%%%%%%%%%%%%%%%%%%%%%%
\begin{figure}[htb]
\includegraphics[width=7.5cm]{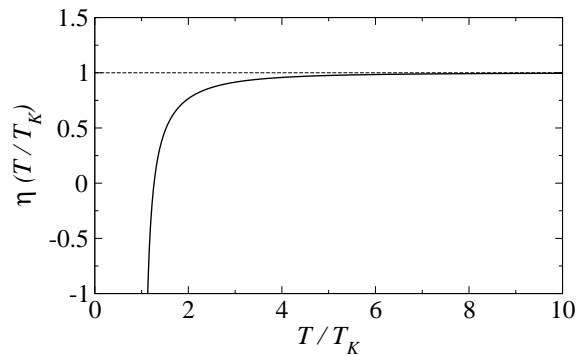}
\caption[]{The renormalization of the $g$ factor due to the Knight
shift. As an estimate, we used $\rho_0 J_\pm=\rho_0J_z=0.15$
\cite{Park}.}
\label{Eta}
\end{figure}
%%%%%%%%%%%%%%%%%%%%%%%%%%%%%%%%%%%%%%%%%%%%%%%%%%%

Let us now study the linear conductance for nonzero bias, $eV =
\mu_{\rm L}-\mu_{\rm R} \neq 0$. We focus on the low-temperature
regime, where we can substitute $-df/d\omega$ by $\left[
\delta(\omega+eV/2) - \delta(\omega-eV/2) \right]/2$ in
Eq.~(\ref{conductance}). We consider the situation where one moves
from the zero points $b_{\bot,0,n}$ to $b_\bot=b_{\bot,0,n} + \Delta
b_\bot$, with $\Delta b_\bot = \eta_{b_{\bot,0,n}} \Delta h_\bot$. If
$|eV| \ll \Delta E_{m,-m}(b_\bot) \ll T_K$, the transmission amplitude
is well approximated by Eq.~(\ref{amplitude}). On the other hand, for
$|eV| \gg T_K \gg \Delta E_{m,-m}(b_\bot)$ the transmission amplitude
is given by
%
%\be 
$\cA_{eV} = J_{\bot,eV}$.
%\ee
%
For the case $|eV|\sim\Delta E_{m,-m}(b_\bot) \gg T_K$ we can expand
$\cA_{\tilde{D}\approx\max\{T,\Delta E_{m,-m}(b_\bot)\}}$ up to second
order in perturbation theory at the flow end, yielding
\bea 
\cA_{\tilde{D}}^{(m)}(\omega)
% & = & J_{\bot,\tilde{D}}^{(m)} +
%\rho_0 \int_{-\tilde{D}+eV/2}^{\tilde{D}-eV/2} d\epsilon'
%\frac{J_{\bot,\tilde{D}}^{(m)}\, J_{z,\tilde{D}}^{(m)}}
%{\omega-\epsilon'} \nn \\
& = & J_{\bot,\tilde{D}}^{(m)} + \rho_0
J_{\bot,\tilde{D}}^{(m)} J_{z,\tilde{D}}^{(m)}\ln \left|
\frac{\omega+\tilde{D}-eV/2} {\omega-\tilde{D}+eV/2} \right|, 
\eea
where the integration limits account for the asymmetric cut of the
bands. Keeping terms up to third order in $J_{\tilde{D}}^{(m)}$ and
combining the results for zero and nonzero bias, we obtain
\be 
\frac{G}{G_0} = \frac{\pi^2\rho_0^2}{16} J_{\bot,\tilde{D}}^2
\left[ \delta_{eV,0} + \rho_0\, J_{z,\tilde{D}} \ln \left(
\frac{\Delta E_{m,-m}}{\left| |eV| - \Delta E_{m,-m} \right|} \right)
\right], 
\ee 
which agrees with ref.~\cite{appelbaum}.
The two split Kondo peaks appear at
$|eV|=\Delta E_{m,-m}(b_\bot)$.
Thus, the distance between the two peaks oscillates with magnetic
field, following the renormalized periodic oscillations of the tunnel
splitting $\Delta E_{m,-m}(b_\bot)$.

These results can be extended to the strong-coupling Kondo regime,
namely at $T=0$, where only the two lowest-lying states
$\left|S\right>$ and $\left|-S\right>$ contribute to the Kondo
effect. Similarly to the spin $S=1/2$ case, our calculations yield
$G(T=0) = \frac{G_0}{2} \sum_s \sin^2\delta_s = G_0 $ at the zero
points of the Berry-phase oscillation, where $|\delta_s| = \pi/2$ is
the scattering phase shift in the unitary limit. In order to find the
zero points of the Berry-phase oscillation, one must employ a more
accurate approach, such as the numerical renormalization group
technique \cite{future}. However, since the ground state of the Kondo
model given by Eq.~(\ref{H_Kondo}) has $S_{\rm Kondo} = S_q-1/2$ due
to the spin screening provided by the itinerant electrons, we conclude
that the spin parity of the SMM effectively changes from even to odd
or from odd to even when one goes from the high- to the
low-temperature Kondo regimes. This means that e.g. Ni$_4$ should
behave as if $S_{\rm Kondo}=7/2$ at $T\ll T_K$.

To help guide the experimental effort on this problem, we provide some
estimates for the Kondo temperature using the expression $T_K = D \exp
\left[ -{\rm arctanh} \left( \sqrt{C^{(S)}}/J_z \right) / 2\rho_0
\sqrt{C^{(S)}} \right]$ derived from Eq. (\ref{eq:flowsol}). Using
$\rho_0 J_z = 0.15$ (similar to ref.~\cite{Liang}) and
setting $\Delta$ equal to the level spacing in the SMM, $D =
|A[S^2-(S-1)^2]| = 9.3$ K, we obtain $T_K = 1.2$ K in Ni$_4$ for the
tunneling between the ground states $m=S=4$ and $m=-S=-4$. The two
crucial ingredients for the experimental observation in SMMs are: (i)
a large spin tunnel splitting and (ii) a large tunneling amplitude
between the leads and the SMM. The first requirement is satisfied by
Ni$_4$. The second one remains an experimental challenge. In the case
of Mn$_{12}$, $\Delta E_{S,-S}(H_\bot\sim 0)\sim 10^{-10}$ K for the
ground state tunneling, which leads to a negligible small Kondo
temperature $T_K$. However, $\Delta E_{4,-4}(H_\bot\sim 0)\sim 0.01$
K, which leads to $T_K\sim 1$ K. Unfortunately, since the excited
levels $m=\pm 4$ are only populated at temperatures of about 1 K, the
levels $m=\pm 4$ cannot be resolved by the electrons in the leads.

In summary, we have shown that the Kondo effect in single-molecule
magnets attached to metallic electrodes is a non-monotonic (possibly
periodic) function of a transverse magnetic field. This behavior is
due to Berry-phase oscillations of the molecule's large spin. The
period of these oscillations is strongly renormalized near the Kondo
temperature and should follow a universal function of temperature that
can be accessed experimentally.
%We propose to measure the Berry phase oscillations through the Kondo
%resonance. This can be done below and above $T_K$.
We argue
%have also shown 
that a newly synthesized family of Ni$_4$ SMMs meets the requirements
for such experiment.

%{\it Acknowledgment}. 
This research was supported in part by the National Science Foundation
under Grants No. PHY-9907949 and No. CCF-0523603. We thank E. del
Barco, L. Glazman, J. Martinek, and C. Ramsey for useful discussions.

%%%%%%%%%%%%%%%%%%%%%%%%%%%%%%%%%%%%%%%%%%%%%%%%%%%%%%%%%%%%%%%%%%%%%%%%%%%%%%

%%%%%%%%%%%%%%%%%%%%%%%%%%%%%%%%%%%%%%%%%%%%%%%%%%%%%%%%%%%%%%%%%%%%%%%%%%%%%%

\end{document}